# An Analytics Approach to Designing Patient Centered Medical Homes

Saeede Ajorlou, Issac Shams, Kai Yang

*Department of Industrial and Systems Engineering, Wayne State University, Detroit, MI 48203*

**Abstract**

Recently the patient-centered medical home (PCMH) model has become a popular team-based approach focused on delivering more streamlined care to patients. In current practices of medical homes, a clinical-based prediction frame is recommended because it can help match the portfolio capacity of PCMH teams with the actual load generated by a set of patients. Without such balances in clinical supply and demand, issues such as excessive under and over utilization of physicians, long waiting time for receiving the appropriate treatment, and non-continuity of care will eliminate many advantages of the medical home strategy. In this paper, by extending the hierarchical generalized linear model to include multivariate responses, we develop a clinical workload prediction model for care portfolio demands in a Bayesian framework. The model allows for heterogeneous variances and unstructured covariance matrices for nested random effects that arise through complex hierarchical care systems. We show that using a multivariate approach substantially enhances the precision of workload predictions at both primary and non-primary care levels. We also demonstrate that care demands depend not only on patient demographics but also on other utilization factors, such as length of stay. Our analyses of a recent data from Veteran Health Administration further indicate that risk adjustment for patient health conditions can considerably improve the prediction power of the model.

# 1  Introduction

Health care delivery is a complex multi-level system in which primary care is the base level and acts as a principal point of consultation for patients. The traditional format of primary care is mainly featured by primary care physicians (PCP), in which each PCP has a designated set of patients, called a patient panel. In current practices of most providers, the panel is simply decided by a predetermined maximum size; that is when the quota is reached, no more patients will be added [1,2]. Typical panel sizes range from 1200 to 1600 patients. However, this number alone cannot reflect the actual health workload generated in the pan-el. For example, a PCP with 1200 young and healthy patients might be generally under-utilized, while one with 1200 elderly patients having multiple comorbidities may experience excessive workload, causing long delays in its panel appointment times and forcing patients to switch their PCPs. It is found that many factors such as patient's age, gender, health status and insurance plan can influence the required healthcare workload. Ostbye and colleagues [3] find that patients with different chronic diseases regularly have different visiting frequencies to their PCPs. Naessens and colleagues [4] discover that the number of chronic conditions in a patient will significantly affect clinical workload and medical cost. Potts and colleagues [5] propose a risk-standard method to adjust the panel size for each PCP calculating disease burden of each physician panel for six chronic diseases. However, there is no description or proof about how the risk values are assigned. Balasubramanian and colleagues [6] apply classification and regression trees (CART) to classify approximately 20,000 patients at the Mayo Clinic of Rochester, Minnesota, into 28 categories by using age and gender as factors, so that each category has different workload patterns.

In recent years, the patient-centered medical home (PCMH) has been introduced as a prominent intervention to improving the US primary care systems with better-quality outcomes at lower costs [7]. This model consists of different health professionals grouped together to provide comprehensive, coordinated, accessible and cost effective care while maintaining high levels of service quality and stability. Each team consists of a group of medical professionals such as primary care provider, registered nurse, nutritionist, social worker, and medical clerk that are well poised to provide many aspects of primary care. Theoretically, medical homes are composed of "joint principles" that ideally complement one another and feed into a comprehensive vision of appropriate primary care delivery. The principles are consisted of having a personal physician with an ongoing relationship, a whole person orientation care for all stages of life, a physician-directed medical practice taking responsibilities for all of the continuing care, a coordinated and/or integrated care system across all elements of the care systems, a continuous emphasis on quality and safety, an enhanced access to care through such systems as open scheduling and expanded hours, and finally an appropriate payment system that recognize the added value provided to PCMH patients [8]. Augmented with modern health information technology, the PCMH is crafted to initiate numerous reforms in health care delivery and reimbursement systems [9].

As of 2007, there was some literature examining the prevalence and effectiveness of medical homes. For instance, Fisher [10] outlined some recommendations for the success of medical homes such as increasing effective communication and sharing of information across health care providers, broadening the medical performance measures to include patients' experience with care and ordinary assessment of outcomes, and establishment of medical-home payment system that share savings among all providers involved. A survey by Commonwealth Fund of 3,535 US adults found that when they were provided with a medical home, racial and ethnic disparities in care access and quality were substantially reduced [11]. Furthermore, having a medical home was associated with more preventive screenings and better man-agement of chronic conditions. The Centers for Medicare & Medicaid Services (CMS) planned to pursue Medicare pilot projects in 400 practices in 8 regional sites, and by 2009, twenty bills promoting the PCMH concept have been successfully introduced in 10 states [12]. Another study within the Group Health system in Seattle showed that a medical home prototype led to 29% fewer emergency visits, 6% fewer hospitalizations, and total savings of $10.30 per patient per month over a twenty-one month period [13]. Bates and Bitton [14] indicated seven health information technology domains deemed to be critical for the success of the PCMH model including telehealth, measurement of quality and efficiency, care transitions, personal health records, and, most importantly, registries, team care, and clinical decision support for chronic diseases. The research also found that access to medical home is associated with lower readmission rates among inpatients [13,15].

Practically, as of December 2009, there were about 26 pilot projects involving medical home being directed in 18 states. These consist of over 14,000 physicians and approximately 5 million patients [16]. Of interest, Veterans Health Administration (VHA) launched a nationwide 3-year program in April 2010 to create PCMHs in more than 900 primary care clinics. Early results indicated dramatic improvements such as reducing the appointment waiting time from as long as 90 days down to one day and decreasing the percentage of inappropriate emergency department visits from 52% to 12% [17].

However, there are difficulties in fully achieving the benefits of the PCMH in practice. Rittenhouse, Shortell [12] point out that much work is needed for the PCMH model to fully leverage the electronic clinical information technology, and to develop new business rules and staffing structures before the anticipated cost savings will take place. From an operations management point of view, one of the key success factors of any health delivery system is achieving a balance between supply and demand of care services. This issue is even more critical for the PCMH model since the clinical supply and demand is *portfolio* (vector) in nature. Unlike health demands, the supply of healthcare services can be treated as deterministic and calculated easily based on head counts and available service hours to be offered from all professional lines on an annual basis. Yet, the estimation of clinical workload portfolio based on key patient and provider attributes is a challenging task and to our knowledge there is no related literature tackling this problem in a team based medical home perspective.

Looking retrospectively at data sets of our US Department of Veteran Affairs (VA) sponsored PCMH project, we observe some structural properties listed as follows:
- Patients (unit of analysis) are grouped within PCMH teams and within VA medical facilities with related patient-level, team-level, and also facility-level predictors. In addition, there exists significant heteroscedasticity within each level of hierarchy.
- The actual health workload (outcome) is captured with two variables, one for primary care and one for non-primary care, and these are correlated at some levels of hierarchy.
- Distributions of both outcomes are not normal at all levels of hierarchy.

In this paper, we develop a multivariate hierarchical based portfolio prediction model that takes into account postulated attributes from different levels such as disease types (patient-level), years of experience of the assigned provider (team-level), and zip-code based distance between the patient's home and his/her assigned facility (facility-level). We also want to propose an intensity score for panel size and staffing level adjustment used at different levels of hierarchy, as it would help decision makers on their PCMH team allocation and budget policy decisions. Finally, we seek to screen highly contributing risk factors to demand portfolio variations, since it would inform program analysts on areas more likely affecting the care portfolio balance.

To the best of our knowledge, our work is the first attempt to develop such a clinical portfolio prediction model for medical homes within the OR/MS community. Our contributions include extending the hierarchical generalized linear model to include multivariate response variables in a Bayesian framework, presenting a Markov chain Monte Carlo algorithm with novel prior specifications to fit the model, and utilizing our proposal on real data from VHA to produce findings that have key public and medical implications. Also our approach allows for passing heterogeneous variances and unstructured covariance matrices for the nested random effects as well as their interactions with responses and covariates simultaneously.

The remainder of this paper is laid out as follows. Section 2 introduces our data sources and study variables. Section 3 describes the main methodology with detailed inference and model fitting strategies. In Section 4 we demonstrate the effectiveness of our proposal with a recent case study from VHA. Some discussion points and future research directions are presented in Section 5.

## 2 Data Source and Study Variables

According to National Center for Veterans Analysis and Statistics (NCVAS), VA operates the largest health care system in the USA with 23 geographically different regions (known as VISNs, or Veterans Integrated Service Networks) separated hierarchically within each VISN by level of care or type into different facilities such as VA medical centers (VAMC), Community Based Outpatient Clinic (CBOC), Vet Center (VC), and so forth. Within each facility, every VA primary care enrollee was assigned to an independent physician or non-physician PCP by a standard process-VA Primary Care Management Module. To ensure sufficient staffing and quality of care, each PCP was appointed a target panel size, taking into account the intensity of primary care visits and availability of resources such as supporting staff and capital.

In this study we collected outpatient data from a random sample of 888 different facilities (which corresponds to 130 VAMCs of all 23 VISNs) during FY11 quarter 3 to FY12 quarter 2. The period of one year is appropriate; according to the VA program professionals, the primary care population at each prac-tice site is not subject to drastic change from one year to the next. The Decision Support System (DSS) and National Patient Care Database (NPCD) files of the VA Corporate Data Warehouse (CDW) were employed to extract demographic, socioeconomic, and other types of variables. In addition, due to its rigorous data validity and availability, we chose DRG (Diagnosis Related Group, 29th version) and its ACC (Aggregated Condition Category) codes for patient case-mix and risk adjustment measures in our predictive analytics [18].

Initially there were 82,000 randomly selected patients with 48 independent attributes coded. All patient visits to primary care and women's health are assembled for a total capture period of one year. Visits from other primary care related clinics, such as Internal Medicine or Geriatric Primary Care, are excluded from the analysis. The two dependent variables are total primary care and non-primary care Relative Value Units (or RVUs), and for each unique SSN, they are calculated by converting the primary care and non-primary care Current Procedural Terminology (or CPT) codes from all patient visits during the fiscal year (according to Centers for Medicare and Medicaid Services model). Simply, the Non-PCRVU refers to all of the non-primary care workload during the year, which could be from one or many visits to outpatient specialty care, and the PCRVU is the primary care workload during the year from outpatient primary care. One advantage of using RVUs in our approach as opposed to simple face-to-face visit counts lies in its ability to further accommodate workloads generated by telephone encounters at the VHA. It is noted that the RVU can be seen as a comparable measure of value for care services used in the US Medicare reimbursement and is determined by assigning weight to factors such as personnel time, level of skill, and sophistication of equipment required to render patient services. The predictor variables include baseline demographic and socioeconomic attributes along with some medical factors such as whether the patient has insurance, to which VA facility the patient has been admitted, and so on. Before presenting descriptives of the independent variables we perform some data-preprocessing activities to prevent unexpected errors during model fitting phase. These include: 1) discarding and imputing (by unconditional mode imputation) missing values of such features as 'VISN' and 'CAN Score' (will be introduced shortly), 2) removing outliers from such variables as 'Age' and 'Assigned provider experience' thus focusing on the first through ninety-ninth percentiles, and 3) binning multimodal, highly skewed features such as 'Distance' and 'Length of stay' into discrete factors. Following this preprocessing, the number of records was reduced to 81190 patients.

To achieve a better picture of the data environment, we tentatively arranged all independent attributes into five groups as summarized in Table 1. It should be noted that these variables remain the same for a patient during the fiscal year. Note that SD stands for standard deviation and % denotes the percentages of the subgroup in the population. 'Priority' levels range from 1-8 and are assigned based on the veteran's severity of service-connected disabilities and VA income means test (VHA Handbook 1601A.03). 'Distance' is calculated in miles between patient's home zip code and the zip code of the facility he/she admit-ted, considering the latitude and longitude of the two locations. Records with a calculated distance greater than 240 miles were excluded and the remaining were converted into three levels. 'Changed provider count' denotes the number of times during the year that the patient changed his/her assigned provider. As mentioned earlier this variable could be a marker of unbalanced workload among PCPs and discontinuity of care received by patients. 'Length of stay' (LOS) displays the number of days spent admitted at a VA hospital. 'CAN Score' is the care assessment need score, which reflects the likelihood of admission or death within a specified time period. This score is commonly expressed as a percentile ranging from 0 (lowest risk) to 99 (highest risk) and it indicates how a VA patient is compared with other patients in terms of the likelihood of hospitalization or death. Each PCMH team has a unique 10-digit code throughout all VA medical systems nationwide. Currently all teams have the same number of professions within all VA centers. The number of PCMH teams and VA facilities in our data set are 6,051 and 287 respectively. ACC categories are determined based on the various ICD-9-CM (International Classification of Disease, ninth version, Clinical Modification) codes assigned to the patient at each visits during the whole fiscal year. They basically indicate the occurrence of a specific disease group, and they are not mutually exclusive categories, meaning that a patient may have more than one ACC during the fiscal year and most actually do.

As shown in the table, the mean age of patients is 62.42 years (SD = 15.26) and about half of the cohort were over age 63 (median = 63). Not surprisingly, near 94% of our veteran population was male and approximately 61% of all were insured. Over half of the patients were married but lower than one third of all were reported as actively employed. The most frequently enrolled patients are the low income and Medicaid group followed by >50% connected disability, and non-service connected patients with income above HUD (Housing and Urban Development). The majority of patients (93%) did not spend a day as an inpatient admitted to the hospital, and most of them travelled only a short distance to receive care from the VA hospitals. The mean care assessment score is roughly 47 with a great variation (SD = 28.88). Also, on average, most of patient's assigned providers are well experienced working rather full time in their roles.

Next, we provide two schematic views of the mean annual care demand and disease *prevalence* of multiple patient groups. In Fig.1, the average RVU demands of the primary and non-primary care generated are displayed across different priority groups with insurance status nested. Not unexpectedly, the non-primary care effort is always more than the primary care workload and its ratio changes from 1.8 in group 8-insured to 6.6 in group 4-uninsured. In all priority groups, uninsured VA patients compared to insured ones produce, on average, more workload in terms of both primary and non-primary care. In addition, the biggest (lowest) workload demands for both primary and specialty care services are associated with group 8-uninsured patients (group 6-insured patients).

Fig.2 displays a mosaic plot of illness types along with patients' gender and their marital status. We excluded ACC 28 (neonate's diseases) and 'unknown' marital category from these analyses because of either the absence or rarity in our sample study. Note that letters P, N, and M above the marital bar denote 'Previously married', 'Never married', and 'Married' groups. The ACC labels are given in the Table 6. As shown, the most commonly occurring conditions among all patient clusters is ACC 30 (Screening) followed by ACCs 5 (nutritional and metabolic) and 16 (heart). However, the least prevalent illnesses among the VA patients are ACC24 (pregnancy-related), ACC13 (developmental disability), and ACC15 (cardio-respiratory arrest). Plus, in almost all disease types, married males are more at risk than two other male groups.

# 3 Methodology

In this section we first propose our approach and present its exclusive properties for modeling the PCMH demand variations. Then we develop a Bayesian framework with novel prior specifications for parameter estimation and model inference.

## 3.1 Model Specification

The PCMH data is hierarchically organized into three nested levels as shown in Fig.3, where patients are grouped within PCMH teams, and teams are in turn nested within VA facilities. Note that PCMH teams are tied to facilities, i.e., a specific team cannot work at different facilities (teams are nested within facilities). Risk factors can be associated with the response variables at each level while patients from the same team (facility) may have more similar outcomes than patients chosen at random from different teams (facilities). For example, we can study the effects of age (patient-level), PCMH assigned provider's experience (team-level), and type of hospital (facility-level) on the outcomes with nested sources of variability. This setting, in addition to health services research, may happen in many other applications such as educational studies where students are nested within schools and successively within school district. It has been shown that ignoring a level of hierarchy in a data can greatly influence the estimated variances and sensitivity [19], can seriously inflate Type I error rates [20], and also can result in errors in interpreting the results of statistical significance tests [21]. As such, multilevel statistical models have been proposed to appropriately analyze the hierarchical (correlated) nesting of data, taking into account the variability associated with each level of the hierarchy [22].

To simplify, we begin by creating a univariate 2-level generalized linear model (GLM) that predicts the primary care RVU (PCRVU) in each PCMH team with one patient-level (age) and one team-level (assigned provider's experience) predictors. The level-1 model would look like

$$y_{ij} = \beta_{0j} + \beta_{1j} X_{ij} + e_{ij} \tag{1}$$

where $y_{ij}$ is the PC workload for patient $i$ in PCMH team $j$ with an exponential family density of form $f(y | x, \phi) = c(y, \phi) \exp\left\{\frac{y\theta - b(\theta)}{\phi}\right\}$, $\beta_{0j}$ is the average PC workload generated in team $j$, $X_{ij}$ is the patient-level predictor (age) for patient $i$ in team $j$, and $\beta_{1j}$ is its coefficient or slope. The parameters $\theta$ and $\phi$ are called canonical (natural) parameter and scale (dispersion) parameter, respectively. Also $c(\cdot)$ and $b(\cdot)$ are determined by the type of (conditional) distribution under study. This way, we assume that each team has a different (varying) intercept coefficient and a different (varying) slope coefficient. These team-specific coefficients can be specified as either fixed effects or random effects [23]. Treating them as fixed effects, however, leads to a large number of parameters with often very poor estimation results. A more conservative way is to think of them as random variables being modeled by some (level-2) *hyperparameters*. The last term, $e_{ij}$, is the patient-level error term which is assumed to be normally distributed with covariance structure R. Unlike most methods in the literature, which suppose that the residual variation is the same at the 2-level (teams) and/or the upper levels of hierarchy, we allow unequal variations of the residual to be passed not only on various levels of the hierarchy but also on different response variables.

The next step is to explain the variation of the (level-1) regression coefficients introducing explanatory variables at the team level like

$$\begin{aligned} \beta_{0j} &= \gamma_{00} + \gamma_{01} Z_j + u_{oj} \\ \beta_{1j} &= \gamma_{10} + \gamma_{11} Z_j + u_{1j} \end{aligned} \tag{2}$$

In this equation, $\gamma_{00}$ is the grand mean of PC workload across patients and across PCMH teams, $\gamma_{10}$ is the average effect of the patient-level predictor (age) across all teams, $Z_j$ is the team-level predictor (assigned provider's experience) for team $j$, $\gamma_{01}$ and $\gamma_{11}$ are its (level-2) intercept and slope regression coefficient, and the $u$-terms are random errors at the team level, which are assumed to be normally distributed with covariance G. Similar to the R-side covariance matrix, we let these level-2 random errors have unequal variances and also leave them free to be correlated with each other. It is worth pointing out that $Z_j$ in the second line of (2) acts as a *moderator* for the relationship between workload and patient age at level-1 analysis; that is, the relationship varies according to the value of the moderator variable. Following the same logic, we can extend this model to add further hierarchies at the facility-level, at the regional level, and so on.

Now a multivariate generalization of this hierarchical GLM is proposed in which both PC and Non-PC workloads are predicted simultaneously. There are several advantages of using a multivariate approach instead of univariate method [24]. One is that the multivariate analysis can better control the type I error rate compared to carrying out a series of univariate statistical tests. Second, this approach can shrink the prediction interval of the dependent variables to a large extent when compared to predicting one of them in isolation. Also using a multivariate scheme, the covariance structure of the responses can be decomposed over the separate levels of hierarchy, which can be of much value for multilevel factor analysis.

Suppose we have $P$ response variables and let $Y_{hijk}$ be the workload on outcome $h$ (PC or Non-PC workload here) of patient $i$ in PCMH team $j$ and facility $k$. Here we put the measures (responses) on the lowest level of hierarchy, and represent the different outcome variables by defining $P$ dummy variables like

$$d_{phijk} = \begin{cases} 1 & p = h \\ 0 & p \neq h \end{cases}.$$

Then we formulate the lowest level as

$$Y_{hijk} = \pi_{1ijk}d_{11ijk} + \pi_{2ijk}d_{22ijk} + \ldots + \pi_{pijk}d_{ppijk} ,$$

in which neither the usual intercept nor the error term exists as before. The reason for this is that we solely serve the lowest level as a way to define the multivariate structure using dummy variables. Then following (1), we may use $\pi$-terms to employ regression equations at the patient level

$$\pi_{pijk} = \beta_{p0jk} + \beta_{p1jk}X_{pijk} + e_{pijk} \qquad (3)$$

in which a separate index is utilized for denoting the dependent variable of interest. It is noted that with this approach one can fit different intercepts and slopes for different response variables and allow them to vary across any levels of hierarchy. Following (2), at the team level, we can have

$$\begin{aligned} \beta_{p0jk} &= \gamma_{p00k} + \gamma_{p01k}Z_{jk} + u_{p0jk} \\ \beta_{p1jk} &= \gamma_{p10k} + \gamma_{p11k}Z_{jk} + u_{p1jk} \end{aligned} , \qquad (4)$$

where we introduce our 2-level predictors (level-1 moderators) along with random intercepts and slopes and finally link them to the facility level equations by

$$\gamma_{p00k} = \lambda_{p000} + \lambda_{p001}W_k + u_{p00k}$$
$$\gamma_{p01k} = \lambda_{p010} + \lambda_{p011}W_k + u_{p01k}$$
$$\gamma_{p10k} = \lambda_{p100} + \lambda_{p101}W_k + u_{p10k}$$
$$\gamma_{p11k} = \lambda_{p110} + \lambda_{p111}W_k + u_{p11k} \quad .$$
(5)

Keeping on this way, one can straightforwardly extend the model to include more predictors at each level and study the effects of fixed and random parameters at any given point. Another advantage of such modeling is that we can impose an equality constraint across all response variables to build a specific relation with certain effects. For example, we can force level-1 regression coefficients for *p*=1 (PC workload) and *p*=2 (Non-PC workload) to be equal by adding the constraint $\beta_{11jk} = \beta_{21jk}$. This makes the new model nested within the original model, and thus we can test whether simplifying the model is justified, using a chi-square test on deviances. Plus, if the predictor has random components attached to it, a similar approach would apply to the random part of the model.

At this point, we specify the structure of random components in the model. As shown, we have two random parts in our method: first is the level-1 residual errors as appear in (3) by $e$-terms, and second relates to (higher level) varying intercepts and slopes introduced by $u$-terms in (4) and (5). We denote the covariance matrix of the former as R and the latter as G and then assume that both are normally distributed with

$$E\begin{bmatrix} \mathbf{u} \\ \mathbf{e} \end{bmatrix} = \begin{bmatrix} \mathbf{0} \\ \mathbf{0} \end{bmatrix}$$
$$\mathrm{Var}\begin{bmatrix} \mathbf{u} \\ \mathbf{e} \end{bmatrix} = \begin{bmatrix} \mathbf{G} & \mathbf{0} \\ \mathbf{0} & \mathbf{R} \end{bmatrix} \quad .$$
(6)

As illustrated, the residual and random parameters are independent having zero means. Generally G and R matrices are large and square with dimensions equal to the number of random coefficients and residuals. While several structures such as spatial or compound symmetry can be thought to formulate those, here we propose an unstructured parameterization tactic by taking the Kronecker product of their decomposed matrices, named **Parametric** and **Structured**, as

$$\mathbf{G} = \begin{bmatrix} \mathbf{P}_1 \otimes \mathbf{S}_1 & \mathbf{0} & \mathbf{0} \\ \mathbf{0} & \mathbf{P}_2 \otimes \mathbf{S}_2 & \mathbf{0} \\ \mathbf{0} & \mathbf{0} & \ddots \end{bmatrix} \quad .$$
(7)

At the moment we focus on G decomposition, but a same logic is applied to R. In (7), $\otimes$ shows the Kronecker (direct) product; **P**-terms represent the Parametric part, which is low dimension and needs to be estimated by data; $\mathbf{S}$-terms stands for Structured part, which is typically high dimensional and assumed as known; and zero-off diagonals express the independence among components. Note that in its simplest case such as general linear models, where the Parametric matrix is reduced to scalars and the Structured part is taken as identity matrices, equation (7) will reduce to the previously known formula $\mathbf{G} = \mathbf{P} \otimes \mathbf{S} = \sigma^2 \mathbf{I}$. Thus we can imply (7) as a generalization for covariance functions of other linear statistical models.

To better describe the structure in (7), we present examples from our case study. Suppose that we are interested to know whether the identity of a VA facility introduces dissimilar amounts of workload variations. Thus we may construct the top left part of (7) like

$$\mathbf{P}_{\text{Facility}} \otimes \mathbf{S}_{\text{Facility}} = \begin{bmatrix} \sigma^2_{\text{PCRVU}} & \sigma_{\text{PCRVU,Non-PCRVU}} \\ \sigma_{\text{Non-PCRVU,PCRVU}} & \sigma^2_{\text{Non-PCRVU}} \end{bmatrix} \otimes \mathbf{I} \qquad (8)$$

which permits heterogeneous variances across workloads (main diagonal) along with their possible correlation (off-diagonal), and further postulates that the facilities are independent to each other (with the identity matrix). So at the worst case for fitting (8), we need 3 degree-of-freedom (DF) to estimate three different elements from the Parametric matrix. Further, we may suspect that it is better to fit age (level-1 predictor) with varying intercept and slopes presented by different teams as

$$\mathbf{P}_{\text{Team}} \otimes \mathbf{S}_{\text{Team}} = \begin{bmatrix} \sigma^2_{\text{(Intercept)}} & \sigma_{\text{(Intercept),Age}} \\ \sigma_{\text{Age,(Intercept)}} & \sigma^2_{\text{Age}} \end{bmatrix} \otimes \mathbf{I} \qquad (9)$$

where the (1:1) element is the amount of variation in regression intercepts among different teams, the (2:2) element is the amount of variation in regression slopes introduced by the patient age across teams, and as before the identity matrix expresses the independence among PCMH teams. Here the model specification is completed and in the next part we explain the model fitting and inference in a Bayesian framework.

## 3.2 Estimation and Inference

Before describing model inferences, we give another but equivalent description of our proposal. By substituting equation (2) into equation (1) and rearranging the terms, we have

$$y_{ij} = \gamma_{00} + \gamma_{10} X_{ij} + \gamma_{01} Z_j + \gamma_{11} X_{ij} Z_j + u_{1j} X_{ij} + u_{oj} + e_{ij} \qquad (10)$$

in which two distinct segments can be implied: the first is $\left[ \gamma_{00} + \gamma_{10}X_{ij} + \gamma_{01}Z_j + \gamma_{11}X_{ij}Z_j \right]$, which we call the deterministic part, and the second is $\left[ u_{1j}X_{ij} + u_{oj} + e_{ij} \right]$, which we call the stochastic part. That way, the moderator effect of (2) is expressed as *cross-level* interaction $X_{ij}Z_j$ and the multiplication $u_{1j}X_{ij}$ directly reveals that the error is different for different values of $X_{ij}$ (heteroscedasticity). Taking a matrix form, we may rewrite the right-hand-side of (10) as $\boldsymbol{\eta} = \mathbf{X}\boldsymbol{\gamma} + \mathbf{W}\boldsymbol{\varepsilon}$, where $\mathbf{X}$ and $\mathbf{W}$ are the design matrices for deterministic and stochastic parts. Then the left-hand-side of (10), conditional on the stochastics, shapes a GLM response of $g\left(E[\mathbf{Y}|\boldsymbol{\varepsilon}]\right)$, where $g(\cdot)$ is a differentiable monotonic link function that allows the outcomes to possess any member of the exponential class of distributions. Now assuming a density function of $q^p(\boldsymbol{\varepsilon}^p; \upsilon^p)$ for the stochastic part of the $p^{\text{th}}$ response variable ($p = 1, 2, \cdots, P$), we can make inferences about the unknown parameters by maximizing the marginal likelihood

$$L(\boldsymbol{\gamma}, \upsilon, \phi | \mathbf{Y}) = \int \prod_{p=1}^{P} f^p(\mathbf{Y}^p | \theta^p, \phi^p) \, q^p(\boldsymbol{\varepsilon}^p; \upsilon^p) \, d\boldsymbol{\varepsilon}^p \quad , \tag{11}$$

where $\boldsymbol{\gamma} = \left[ \boldsymbol{\gamma}^1, \boldsymbol{\gamma}^2, \cdots, \boldsymbol{\gamma}^P \right]$ is the vector of deterministic coefficients, $q^p(\boldsymbol{\varepsilon}^p; \upsilon^p)$ is a multivariate Gaussian distribution of dimension $P$ with mean zero and variance-covariance $\upsilon^p$, and $\phi^p$ and $\theta^p$ are the GLM scale and canonical parameters, respectively.

Generally two basic methodologies have been expressed in the literature for optimizing a univariate version of (11): the first one tries to approximate the model based on linearization and pseudo-data with fewer nonlinear components, such as the pseudo-likelihood technique of Wolfinger, O'connell [25]. The second category consists of integral approximation methods that attempt to approximate the log likelihood of (11), such as adaptive Gaussian quadrature [26]. But both approaches have some key drawbacks that, we think, cause them inappropriate for our study context. For example, a true objective function for the overall optimization does not exist in the first class; thus it potentially produces estimates that are inconsistent under standard (small domain) asymptotic assumptions. Additionally, the bias size can be substantial in the case of major variance components or few observations per participant. Similarly, methods in the second approach cannot accommodate R-side covariance structure such as overdispersion parameter [27]. These problems also become more crucial when more than one outcome needs to be estimated [28].

Due to this, we decide to put forward a Bayesian framework that utilizes an exact maximum likelihood approach by numerical integration techniques. To this end, we need to first determine suitable priors for the parameters of interest then employ a simulation-based integration technique, such as Metropolis-Hastings or slice sampling, to iteratively sample the posterior until convergence. Afterwards, generated samples are used to estimate the approximate expectations of quantities of interest. However, setting up the appropriate priors can greatly affect inference about posteriors, because in many cases, *diffuse* priors and/or *improper* priors lead to improper posteriors upon which no valid inference can be made [29]. Accordingly, for the deterministic coefficient vector $\gamma^p$, we use a Gaussian prior of form $N(\gamma_0, \Gamma)$. Moreover, to sample from $\eta$, since its distribution cannot be identified, we apply the Metropolis-Hastings update of Damlen et al. [30]. In summary, the method is updating $\eta$ in some blocks; each consists of groups of residuals expected to have some form of residual co-variation as defined by the R structure. That way, the conditional density of $\eta^p$ is formulated as

$$f(\eta_l^p \mid \mathbf{Y}^p; \gamma^p, \varepsilon^p) \propto \prod_{i \in l} p_i(\mathbf{Y}_i^p \mid \eta_i^p) f_N^p(\mathbf{e}_l \mid \mathbf{0}, \mathbf{R}_l) \tag{12}$$

where *l* stands for blocks of $\eta^p$ with non-zero residual covariances, $f_N^p$ indicates a conditional multivariate normal distribution for the linear predictor residuals, and $p_i(\mathbf{Y}_i^p \mid \eta_i^p)$ is the probability of data point $\mathbf{Y}_i^p$ (from $p^{\text{th}}$ outcome) with linear predictor $\eta_l^p$.

In order to update the parameter vector $\rho^p = [\gamma^T, \varepsilon^T]^T$, the single-block Gibbs sampler of García-Cortés, Sorensen [31] is applied. Essentially, the method solves the sparse linear system of $\tilde{\rho}^p = \mathbf{A}^{-1}\mathbf{M}^T\mathbf{R}^{-1}(1 - \mathbf{M}\rho_*^p - \mathbf{e}_*^p)$ using Cholesky decomposition technique. In the formula, $\mathbf{A}$ is the coefficient matrix of form $\mathbf{A} = \mathbf{M}^T\mathbf{R}^{-1}\mathbf{M} + \begin{bmatrix} \Gamma^{-1} & \mathbf{0} \\ \mathbf{0} & \mathbf{G}^{-1} \end{bmatrix}$, in which $\mathbf{M} = [\mathbf{X}\ \mathbf{W}]$ is the whole design matrix, $\Gamma$ is the prior (co)variance matrix for the deterministic part, and $\{\rho_*^p, \mathbf{e}_*^p\}$ are random realizations drawn from multivariate normal distributions $\rho_*^p \sim N\left(\begin{bmatrix} \gamma_0 \\ \mathbf{0} \end{bmatrix}, \begin{bmatrix} \Gamma & \mathbf{0} \\ \mathbf{0} & \mathbf{G} \end{bmatrix}\right)$ and $\mathbf{e}_*^p \sim N(\mathbf{M}\rho_*^p, \mathbf{R})$ respectively. Based on these, the desired prior sample of $f(\rho^p \mid \eta^p; \mathbf{M}, \mathbf{R}, \mathbf{G})$ is given by $\tilde{\rho}^p + \rho_*^p$.

For taking samples of the variance structures $\mathbf{R}$ and $\mathbf{G}$, we need the sum of squares matrix associated with each diagonal component of (7). This is given by $\mathbf{H} = \mathbf{\Phi}^T\mathbf{S}^{-1}\mathbf{\Phi}$, where $\mathbf{\Phi}$ is a stochastic matrix in which each column is related to the relevant row/column of Parameteric matrix $\mathbf{P}$ and each row is associated with the related row/column of Structured matrix $\mathbf{S}$. In this way, $\mathbf{P}$ can be Gibbs sampled in one block from the *Inverse-Wishart* (IW) distribution $\mathbf{P} \sim IW((\mathbf{H}_p + \mathbf{H})^{-1}, n_p + n_\Phi)$, where $n_\Phi$ is the number of rows in $\mathbf{\Phi}$, $\mathbf{H}_p$ is the prior sum of squares, and $n_p$ is its degrees of freedom. It should be noted that IW is a conjugate prior for the covariance matrix of a multivariate normal distribution.

Usually the goodness-of-fit of Bayesian models can be assessed using the deviance information criterion (DIC), which is a Bayesian alternative to AIC and Schwarz criterion. The DIC can be calculated at different levels of hierarchy and a smaller amount indicates a better fit to the data while compensating for model complexity. Here, we adopt the method of Spiegelhalter et al. [32] and define the *deviance* as $D = -2\log(\Pr(\mathbf{Y}/\mathbf{\Omega}))$, where $\mathbf{\Omega}$ are some parameters of the model. We calculate this probability for the lowest level of the hierarchy at each iteration. In the formula, in case of Gaussian responses we have $\mathbf{\Omega} = \{\boldsymbol{\rho}, \mathbf{R}\}$ and the likelihood would be the normal density $f_N(\mathbf{Y} | \mathbf{X}\boldsymbol{\gamma} + \mathbf{W}\boldsymbol{\varepsilon}, \mathbf{R})$. On the other hand, when the responses are not normal, $\mathbf{\Omega} = \boldsymbol{\eta}$ and the likelihood would change to $\prod_i f_i(\mathbf{Y}_i^p | \boldsymbol{\eta}_i^p)$, where the argument denotes the conditional probability of the $i^{\text{th}}$ data point (lowest level of hierarchy). In other words, for non-Gaussian responses, deviance is obtained by the probability of the data given the linear predictor $\boldsymbol{\eta}$, whereas in normal responses, it is calculated using the probability of the data given the parameters. The DIC can then be attained by $\text{DIC} = 2\bar{D} - D(\bar{\mathbf{\Omega}})$, where $\bar{D}$ is the mean deviance of all iterations and $D(\bar{\mathbf{\Omega}})$ is the deviance evaluated at the mean estimates of the parameters.

## 4 Analytics

### 4.1 Model Fitting and Diagnostics

We conduct multiple analyses to estimate the effect of different patient factors such as disease types (ACCs) on the mean annual primary and non-primary care. To employ our method we first determine the appropriate distributions for the two responses. Here the standard Quantile-Quantile plot along with maximum likelihood method is used, but one can also employ non-parametric techniques such as kernel density estimation. We examine different base densities such as gamma, lognormal, beta, and Cauchy, then judge the best choice as having the best graphical pattern in QQ plot and the biggest likelihood value simultaneously. Based on these criteria, the lognormal distribution is found the most proper case for both RVUs. Fig. 4 shows the QQ plots along with bootstrapped point-wise confidence envelopes at 0.95 accuracy rate. As shown, the PCRVU (left panel) displays a perfect linear pattern, and even for Non-PCRVU (right panel), almost all points lie within the confidence band. We also get the minimum value of the minus log-likelihood based on ML fitting when the lognormal distribution is taken.

To determine the appropriate link function $g(\cdot)$, a range of classical options including log link and inverse link are evaluated by two goodness-of-fit measures, namely DIC and modified HL test [33]. Based on the results (not shown here) we observe that the (default) identity link does estimate the upper and lower tails of both RVUs, accounted for the covariates, more properly than other links, and thus it is chosen for our study.

Since failing to specify the suitable probability density for priors can result in inferential and numerical problems as discussed in Section 3.2, for the deterministic parameters we pick a multivariate normal density with zero vector for the mean $\gamma_0$ and a diagonal matrix of large variances (1e+10) for $\Gamma$. This way we can make sure that the prior is always *proper*. However, for each (decomposed) *block* of the G-(R-) side, we are required to specify the hyperparameters through the IW distribution, which takes two scalars; the expected (co)variance at the limit and the degree of belief parameter. We configure several prior specifications not only for these two parameters but also for different shapes (degrees of freedom) the decomposed matrices can take, then assess the impacts on the DIC measure and their posterior distributions (with MCMC diagnostics). A few such comparisons are discussed in Section 4.2, but now for the first step of our modeling strategy (discussed later), we choose a diagonal matrix of 1/3 (similar to (8)) for all three hierarchies (patients, PCMH teams, and facilities) with 2 degrees of freedom. Scaling outcomes to have a unit variance before the analysis, this prior implies that the total variance is equally split across all three levels together with *a priori* independence of PC and Non-PC workloads.

Although different modeling strategies could be selected for estimating our multilevel model, we focus on the most parsimonious and best-fitting approach for the given data and our specific research questions. To this end, six models (Table 2) from basic to comprehensive are run sequentially and the outputs are reported for each step in order to provide insights for a particular objective. Further, to avoid overfitting within each step, we perform stepwise selection for the deterministic covariates with probabilities to enter and stay of 0.15 and 0.1 respectively. Alternatively, one can employ a Bayesian selection to determine a variable subset [34]. Different functional forms of covariates, such as logarithmic and power relations, as well as within-level interactions are evaluated too at each step but only the statistically significant ones are included. As an example, we analyze 12 pairs of ACC interactions that are notable for co-occurring in patients with multiple chronic illnesses and/or an acute disease combined with a chronic condition [35].

The improvement in model fit is evaluated by DIC (see Section 3.2) over all iterations after the burn-in phase of MCMC simulations. Based on a rule of thumb, we favor the model with lower DIC when the DIC reduction of more than 10 units is observed. Depending on the goodness-of-fit and significance tests, sometimes intermediate models, such as a reduced version of model 3 with only one signifi-cant random slope, are also examined. Performing this strategy, we seek to answer the following three research questions:

- How much of the variance in PC and Non-PC workload is associated with patients, PCMH teams, and VA facilities?
- Does the effect of any patient-level predictor change among PCMH teams or VA facilities? And does the effect of any team-level predictor vary among VA facilities?
- What is the impact of patient *non-adherence* (as measured by "Changed provider count") on PC workload, controlling for patient, PCMH team, and VA facility characteristics?

Setting the significance level at 0.05, we run the models with 50,000 iterations, a burn-in period of 10,000, and a thinning interval of 25. All analyses and computations are done in R version 3.0.2 [36]. In order to address the first question, we fit the unconditional model as summarized in Tables 3-5. Note that the first (third) row in each table shows PC (Non PC) intercept variance along with its 95% Highest Posterior Density interval, and the second row corresponds to the workload correlations. The team ICC for the PC outcome is computed as $\frac{0.168}{0.609+0.168+0.218}$. Simply put, we find that about 17% of the variation in PC workload exists between PCMH teams and 22% is there between VA facilities, leaving near 61% of the variance to be accounted for by patients. Thus a practically meaningful proportion of all variation happens at higher levels, providing support for our use of a 3-level hierarchical model. These percentages are 5%, 16%, and 79% for Non-PC workload respectively. Other useful points can be made by interpreting the correlations among PC and Non-PC at different levels. First, the results of a joint conditional independence test Gueorguieva [37] show that the RVUs (at the patient level) are positively associated which confirms the fact that a simultaneous modeling of both primary and non-primary care is more reasonable than using one of them in isolation. Second, we infer that the correlation is not significant when it comes to the team level, and it is poorly significant at the facility level. Hence, we re-parameterize structure (8) in a way that the off-diagonal is replaced by zero. By doing so, we save two DF, but the changes in variance estimates are too trivial to restate here.

We continue our modeling effort to include predictors and random components at all levels, and then answer other research questions based on the outputs from the best fitting model. For brevity we will not walk through all detailed outputs at each stage, and instead summarize them in Table 6. Also note that level-2 and level-3 predictors are displayed italic. In each row, the first number is for PC and the second is for Non-PC outcome, with ('), ("), () displaying significance at 0.05, <0.001, and non-significance respectively. It worth noting that we suppress the overall intercept since otherwise, the parameter estimates associated with PC are translated as contrasts with Non-PC. Also for team-level, facility-level, and interactions, we only include those factors that are significant in at least one of the six models.

Graphically assessing the relation of age with the outcomes, we observe that both responses have a sigmoidal trend at team levels thus we decide to fit its nested random components with covariance matrix like $\mathbf{P}_{\text{Team}} \otimes \mathbf{S}_{\text{Team}} = \begin{bmatrix} \sigma^2_{\text{(Intercept)}} & 0 & 0 \\ 0 & \sigma^2_{\text{Age}} & 0 \\ 0 & 0 & \sigma^2_{\text{Age}^2} \end{bmatrix} \otimes \mathbf{I}$. A similar structure is fitted for CAN Score as well, but with square root instead of second power relation. For 'Changed provide count' we first test a structure with both random intercept and slope at the facility-level, but after failing to reject the null hypothesis of intercept, we reduce it to random slope only. For fitting insurance covariance, again we first try $\begin{bmatrix} \sigma^2_{\text{Insured}} & \sigma_{\text{Insured,Un-insured}} \\ \sigma_{\text{Un-insured,Insured}} & \sigma^2_{\text{Un-insured}} \end{bmatrix}$, and then drop the correlation after the significance test.

According to the DIC index shown at the bottom of Table 6, we realize that each forward model exhibits a better fit to the data, so we take model 6 to answer the remaining research questions. In order to further validate the final model, we apply model 6 to FY12 quarter 3 data and find almost identical results. We repeat the joint independence test of Gueorguieva [37] for model 6 and reaffirm the positive correlation of responses at the patient level. Put differently, we find that after controlling for all sources of variation, if the primary care workload is increased from one patient to another, on average we will expect an increase in the related non-primary care. In the table, the estimates for deterministic effects are interpreted as prevalence ratios but variance components are reported in natural scale. Also note that the data is scaled to have a unit variance before analysis.

It is worth to highlight that some estimates are changed in terms of significance among models. For example, age, insurance, and CAN Score are significant in Model 2 but no longer significant in later models once their related random slopes are introduced in Model 3. Examining other random components in these models, we figure out that significant variability exists in their nested random intercepts and slopes, even after controlling for these patient-level predictors. Hence, we can say that the association between these variables and the outcomes varies considerably among PCMH teams. Thus we expect that the influence of patient oldness on care demands may be stronger or weaker from one PCMH team to another within a VA facility. The same thing happens in terms of effect magnitude for 'Changed provider count' between Models 4 and 5; the relationship between this variable and both workloads changes meaningfully among different VA facilities. By these statements, we tackle our second research question.

To answer the last research question, we look at the deterministic effect of 'Changed provider count' in Model 6. As shown, for each time that a patient switches assigned provider, we will expect an average of 6% more workload in his/her primary care, after accounting for variations of his/her non-primary care demands. Other selected key findings from Model 6 can be summarized as below:

- Adjusting for the contributions of all other variables, female VA patients tend for produce about 57% more PC (98% more Non-PC) compared to males. This is not unexpected due to gender imbalance issue existed in VA patients.
- Inpatient cohort generally creates 28% more workload in non-primary care compared to outpatients, after accounting for variations of their primary cares.
- Catastrophically ill veterans (P4) have 1.15 times the Non-PC demands of the P8 comparison group. The increase rates are about 35% and 23% for veterans exposed to Agent Orange (or other herbicides) and >50% for disabled veterans. Having been exposed to such chemicals also notably affects the increased caress for cardio-respiratory arrest.
- Change rates in primary cares range from 7% decrease for ACC29 (Transplant) to 52% increase for ACC4 (Diabetes). For non-primary cares, this varies from 11% reduction for ACC11 (Substance Abuse) to 99% rise for ACC30 (Screening).
- Both team-level (patient non-adherence) and facility-level (distance) predictors are significantly associated with the outcomes: Patients travelling more miles to VA hospitals are likely to generate a larger amount of care than closely located patients.
- In co-occurring diseases studies, diabetes greatly interacts with some acute and chronic conditions. For instance, in patients with cardio-respiratory arrest, having diabetes is associated with a 13% (14%) increase in primary care (Non-PC) workload. Another comorbid condition that poses a similar pattern is heart disease, especially for cerebrovascular patients.
- Risk adjustment for disease types and their interactions improves the model fit to a great extent (about 160K reduction in DIC) and makes most of their related effects statistically significant.

Now we present some diagnostic tests for verifying the accuracy of Model 6. First, to assess the Markov chain convergence and mixing properties, trace plots and smoothed posterior densities are provided for each parameter of interest. As an illustration, Fig.5 shows the plots for age and gender across both outcomes and Fig.6 displays them for R-side covariance components. As depicted in Fig.5, the traces are trendless and the chains are mixing well travelling quickly to the target distribution with small autocorrelations.

Nearly same patterns are observed in Fig.6, but chains are now mixing marginally at a bit slower traverse rate, which can easily be tackled by increasing the MCMC iterations. Nonetheless, the densities do smoothly estimate the mean posterior for residual variances as reported in Table 6. For deterministic terms in Fig.5, however, the posterior histogram is plotted in log scale. We additionally perform Gelman, Rubin [38] and effective sample size tests to all posterior estimates (not shown here) and no violations are found therein.

After additional validation steps such as Copas [39] test of overfitting and posterior correlation diagnostics of estimated parameters, we develop two operational indices, namely, hospital level normalized intensity score (NIS) and hospital level risk-standardized utilization rate (RSUR) as

$$NIS_j = \frac{\text{Sum of the total predicted workload of all patients at the given facility}}{\text{Number of patients in the facility} \times \text{Median predicted workload}}$$

and

$$RSUR_j = \frac{\text{Sum of the total predicted workload of all patients with random components}}{\text{Sum of the total predicted workload of all patients without random components}}.$$

The NIS can be used to adjust the panel size up or down for a given hospital, or even for a specific PCMH team within a hospital. Note that the random components are implicitly included in the formula. On the other hand, RSUR indicates the ratio of *predicted* (technically called shrinkage estimate) to *expected* utilization; the numerator computes the PC/Non-PC workload when patients are treated as the specific hospital and denominator calculates the workload as if patients are treated at a so called 'reference' (or normal) hospital. Thus values greater than one reveals that the hospital is over-utilized as compared with the national average range.

## 4.2 Numerical Comparisons

In this section we design three comparison studies to demonstrate some novel aspects of our proposal. First, we evaluate an alternative variance structure with the one applied in Model 6 in terms of the goodness-of-fit measure. Particularly, for patient (residual), team, and facility random intercepts in scenario (1), we change the Parametric matrix to have the same diagonal elements with zero off-diagonals then compare the results with the structure used in Model 6. We run each model twice to take control of the Monte Carlo error and keep all other factors constant among different fittings. As shown in Table 7, the best fit is corresponding to the first row in which the proposed variance structure is applied at all levels of hierarchy.

Second, we investigate the impacts of the random component's prior specification on MCMC diagnostics and posterior distributions. To this end, the DF is kept fixed, and then two alternatives for the expected limit (co)variance { 1.One 2.REML estimate of Wolfinger, O'connell [25]}, as well as other values for the IW degree of belief { 0.002, 0.02, 0.2, and 1 } are assessed. The values used in Model 6 for these two are 1/3 and zero. Results (available from authors) denote that almost no change occurs in deterministic estimates, DIC measure, and directions of (co)variance components. However, the absolute range of alternations in variance estimates is around 2.3% that the base values in Model 6. We detect that better chain convergence and mixing property is observed when using priors with smaller limit (co)variances and larger (near one) degree of the belief parameter. Further, the posterior correlation estimates remain reasonably unchanged while examining different types of priors, which provide some reassurance that our priors do not dominate the model to an unacceptable extent.

Lastly, we perform comparisons between our proposal and the situations when one employs a series of univariate (multilevel) GLMs for predicting the outcomes. To this end, we keep Model 6 settings constant and consider two scenarios: 1) A bivariate 3-level GLM with joint primary and non-primary care workloads, and 2) Two univariate 3-level GLMs one for primary care (PC) and one for non-primary care (Non-PC) workload predictions. Fitting both models, we aggregate the credible intervals for the mean outcomes and then compare them with the actual values. Interestingly, the probability of *joint* correct prediction (for both responses) is about 67% for the first scenario and about 58% for the second. Then we pick those correct intervals, compute their ranges {max–min}, and calculate basic statistics for the ranges in Table 8. As displayed, the credible intervals are substantially narrowed when applying the multivariate approach. Thus we can conclude that a joint modeling of primary and non-primary care workloads would provide more robust and realistic predictions for medical home practices.

## 5 Discussion

A key factor in the success of medical homes in delivering quality and coordinated care lies in their teams' ability to handle uncertainties that can be caused by different sources such as patient/physician appointment scheduling, care logistics, and more importantly patients' health demands. This paper addresses the problem of clinical demand prediction in the presence of nested sources of variation at different operational levels. We collected outpatient visit data from a large sample of Veterans Affairs hospitals and investigated the relationship between risk factors at three operational levels and total care demands on a yearly basis. We propose a multivariate multilevel generalized linear model in a Bayesian framework to predict the care demand portfolio in medical home practices. The proposal can fit heteroscedastic variances and unstructured covariance matrices for nested random effects and residuals as well as their interactions with categorical and continuous covariates simultaneously.

We find that utilizing a multilevel analysis with nested random components can greatly contribute to model fit in hierarchical healthcare systems. Further, we show that risk-adjustment for patient disease conditions and their comorbidities extensively enhances the prediction power of our model. Our results confirm that using a multivariate as opposed to a univariate approach can significantly shrink the correct credible intervals for workload predictions thus allowing for a more precise estimation of either outcome. The approach used in this paper has a general application and could also be employed for analysis of multiple health outcomes in a variety of health analytics contexts.

Turning to specific results from recent VA data, we see that overall, the primary care is positively associated with the non-primary care (correlation coefficient of 0.027 as appeared in Fig.6) after accounting for all studied sources of variability. We find the association between patient-level predictors such as age and the care workloads varies considerably among PCMH teams within a hospital. Further, the effect of patient non-adherence on care demands is subject to change from one hospital to another. Moreover, it is found that patient oldness can contribute to the increased care demands required for heart, nutritional, and gastrointestinal diseases.

Finally there are some limitations to this research that need to be mentioned. First, the data in our study are collected solely from a veteran population (with fewer female and more senior patients) who receives support from government budgets. Thus the results from our study may not fully generalize to other health care systems. Second the data used is administrative and not real time, so some issues such as model tuning and calibration should be taken into account when dealing with online prediction efforts.

Our work can further be extended in some fronts. One challenging direction would be to modify the proposed approach to handle longitudinal observations from past history of care demands for a specific patient profile. This may be done by expanding the multivariate distribution of outcomes to include a temporal dimension which requires great care in model specification and implementations thanks to various inter-correlations. Alternatively, one can combine some autoregressive terms to the variance structure introduced in this work. Another issue worth exploring is related to the way that one can adjust for patient risk or comorbidities. Although several algorithms such as Clinical Risk Group (CRG), *veriskhealth* DxCG, and CMS's HCC software have been used in the literature, no scientific study is available to systematically evaluate the impacts of each algorithm on prediction modeling of care demands.

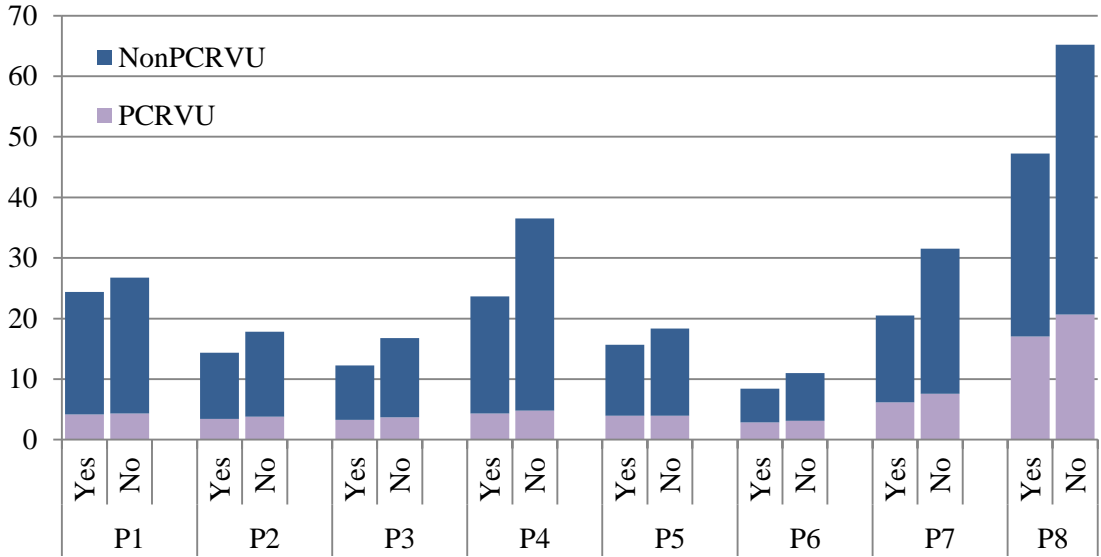

Fig.1: Average annual RVU demands across patient priority and insurance status

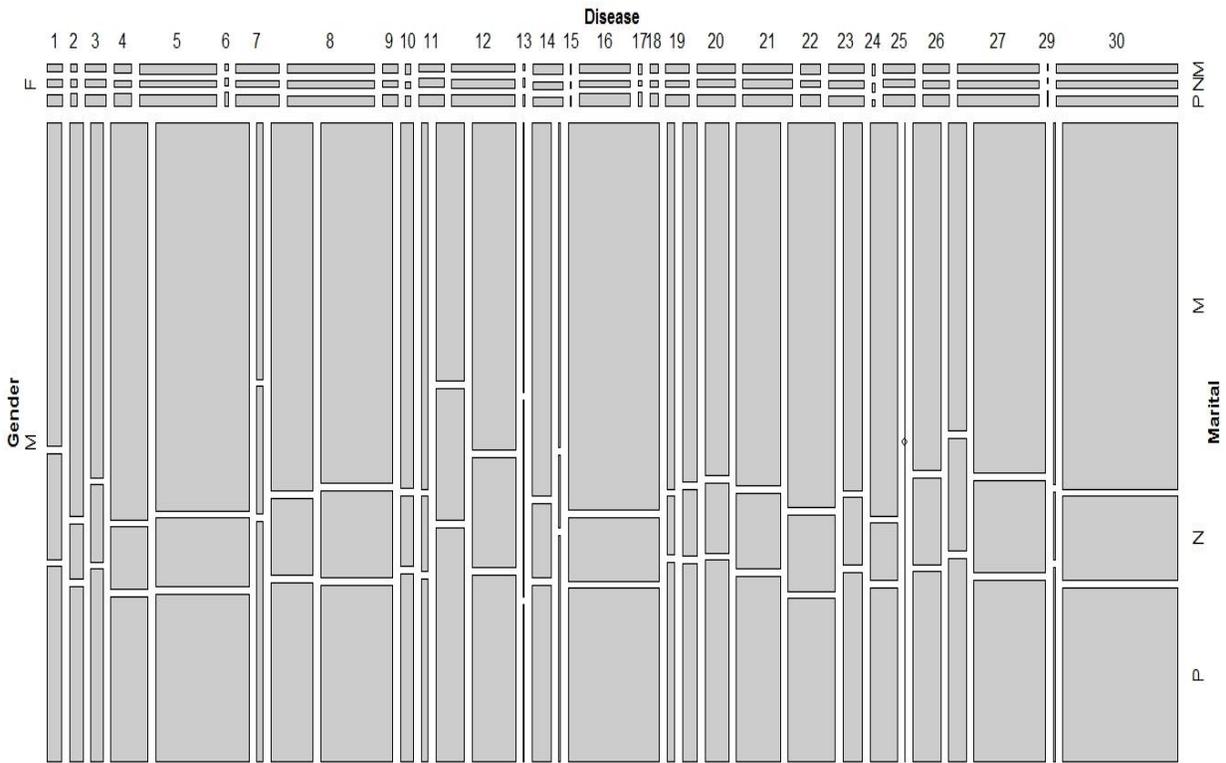

Fig.2: Mosaic plot of disease prevalence across patient gender and marital status

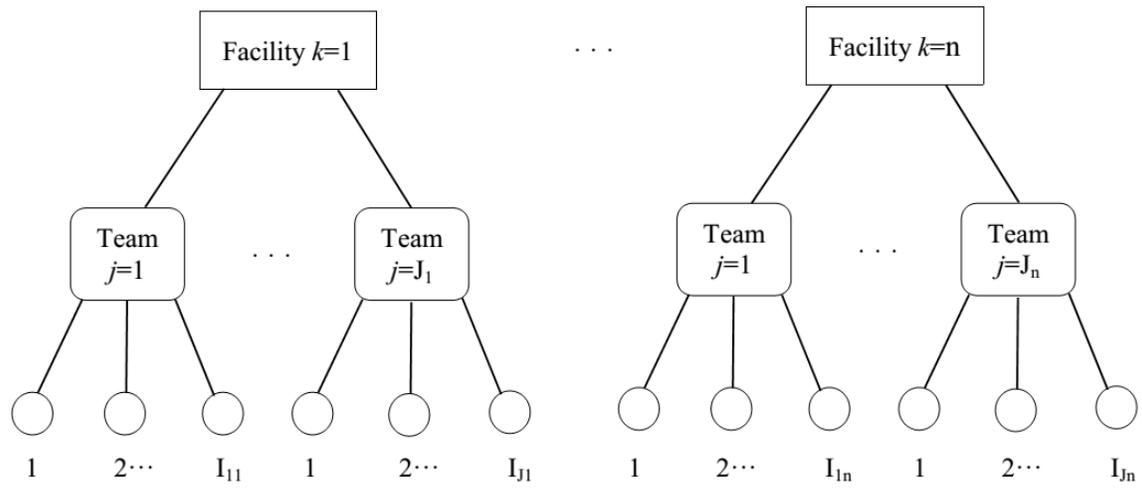

Fig.3: Data structure for PCMH hierarchical model

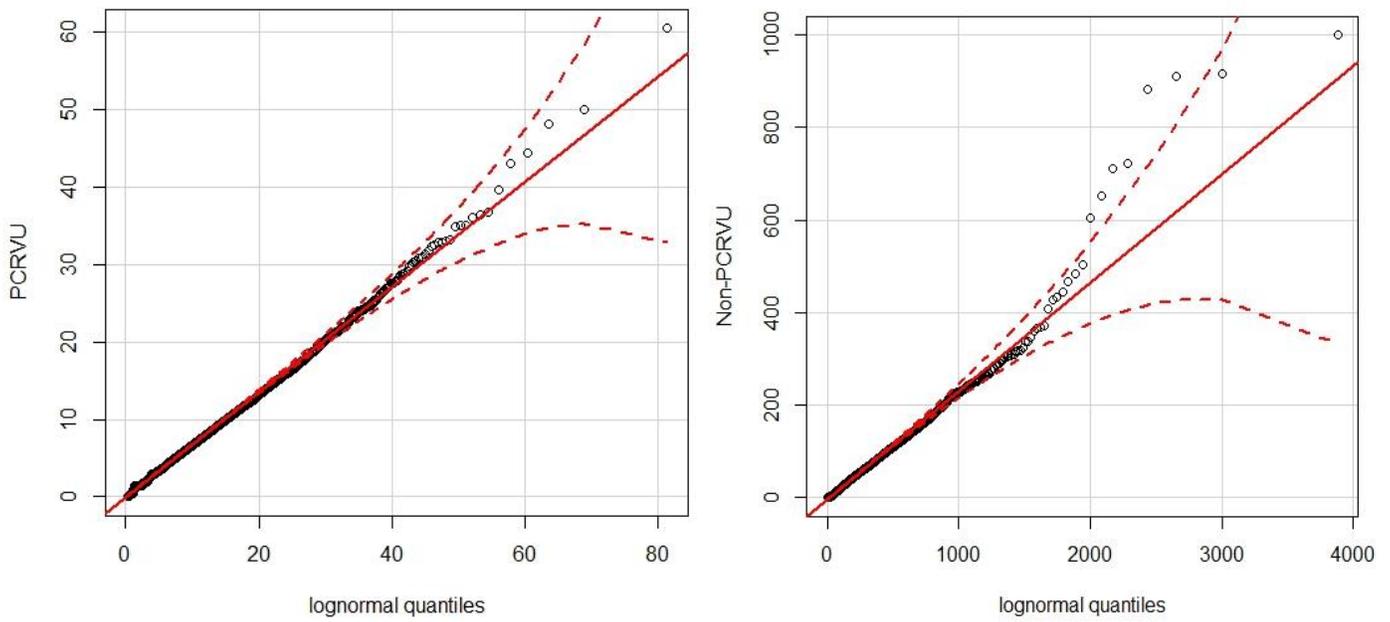

Fig.4: QQ Plots of PCRVU (left) and Non-PCRVU (right) with 95% confidence bands

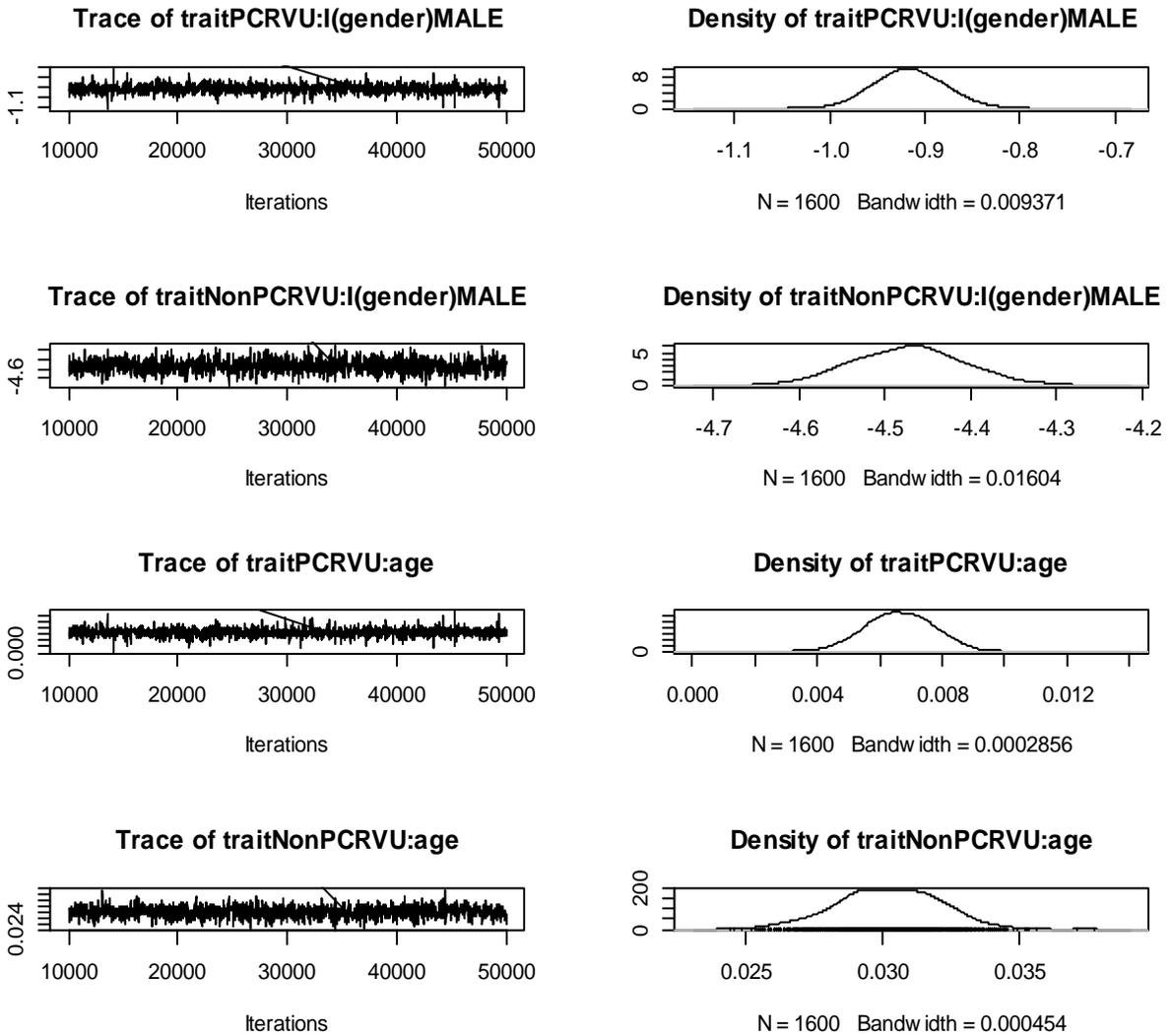

Fig.5: Trace plot and posterior density estimates for the effects of 'age' and 'gender' on primary and non-primary care relative value units

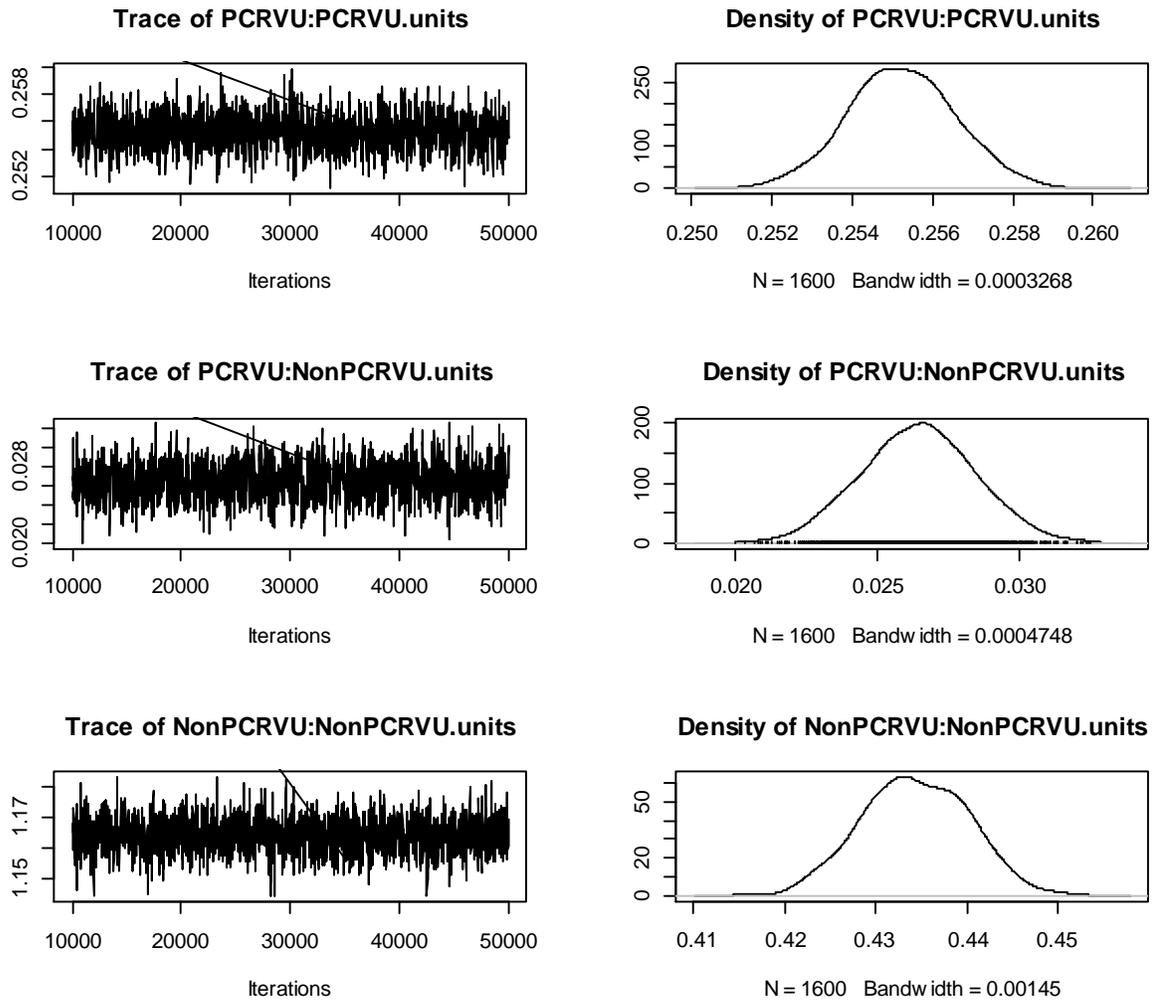

Fig.6: Trace plot and density estimates for (co)variance components of primary and non-primary care relative value units at patient (unit) level

Table 1: Baseline characteristics of patient factors ($n = 81190$)

| Group | Attribute | Mean (SD) | $n$ (%) |
|---|---|---|---|
| Demographic | Gender | | |
| | Male | | 76247 (93.91) |
| | Female | | 4943 (6.09) |
| | Age (as of 7/1/2011, years) | 62.42 (15.23) | |
| | Marital status | | |
| | Married | | 46634 (57.44) |
| | Previously married | | 22520 (27.74) |
| | Never married | | 11559 (14.24) |
| | Unknown | | 477 (0.58) |
| Socioeconomic | Insurance (of any types) | | |
| | Yes | | 49551 (61.03) |
| | No | | 31639 (38.97) |
| | Employment status | | |
| | Active Military Service | | 134 (0.16) |
| | Employed Full-Time | | 17008 (20.95) |
| | Employed Part-Time | | 4013 (4.94) |
| | Not Employed | | 28619 (35.25) |
| | Retired | | 28517 (35.12) |
| | Self Employed | | 2039 (2.51) |
| | Unknown | | 860 (1.07) |
| Enrollment | Priority | | |
| | 1 (service connected disability > 50%) | | 18404 (22.67) |
| | 2 (service connected disability 30%–40%) | | 6548 (8.07) |
| | 3 (service connected disability 20–30%) | | 9859 (12.14) |
| | 4 (catastrophically disabled) | | 2285 (2.82) |
| | 5 (low income or Medicaid) | | 21258 (26.18) |
| | 6 (Agent Orange or Gulf War illness) | | 3697 (4.55) |
| | 7 (non-service connected, income below HUD) | | 2243 (2.76) |
| | 8 (non-service connected, income above HUD) | | 16896 (20.81) |

Table 1— continued: Baseline characteristics of patient factors ($n = 81190$)

| Group | Attribute | Mean (SD) | $n$ (%) |
|---|---|---|---|
| Utilization | VISN | | |
| | 1 (New England Health Care System) | | 3371 (4.15) |
| | 2 (Network Upstate New York) | | 1846 (2.27) |
| | … | | |
| | Facility | | |
| | 662 (San Francisco) | | 109 (0.13) |
| | 537 (Chicago) | | 235 (0.29) |
| | … | | |
| | Distance | | |
| | Near (Below 25 miles) | | 61430 (75.66) |
| | Middle (Between 25 and 50 miles) | | 12344 (15.21) |
| | Far (Above 50 miles) | | 7416 (9.13) |
| | PCMH team | | |
| | 1000001805 | | 13 (0.02) |
| | … | | |
| | Assigned provider position | | |
| | Primary care physician | | 55470 (68.32) |
| | Assistant physician | | 5174 (6.37) |
| | Attending physician | | 7365 (9.07) |
| | Nurse practitioner | | 13181 (16.24) |
| | Assigned provider experience (years) | 8.40 (7.77) | |
| | Changed provider count | 0.74 (0.89) | |
| | Provider full time equivalent | 0.86 (0.24) | |
| | Length of stay (inpatient-day) | | |
| | Zero | | 75522 (93.02) |
| | Non-zero | | 5668 (6.98) |
| Clinical | CAN Score | 47.32 (28.87) | |
| | ACC | | |
| | 1 (infectious and parasitic) | | 9181 (11.31) |
| | 2 (malignant neoplasm) | | 7940 (9.78) |
| | … | | |

Table 2: Regression modeling strategy and specific results for 3-level hierarchical model

| Model 1 | Model 2 | Model 3 | Model 4 | Model 5 | Model 6 |
| --- | --- | --- | --- | --- | --- |
| No predictors, just residual and random intercepts (Unconditional) | Model 1 + patient-level predictors | Model 2 + random slopes for patient-level predictors | Model 3 + team-level predictors | Model 4 + random slopes for team-level predictors | Model 5 + facility-level predictors |
| Results used to compute Inter-class Correlation Coefficient (ICC) which assesses the degree of clustering among subsets of cases in the data. | Results show the relationships between patient-level predictors and outcomes | Model 2 results + findings that show if the associations between patient-level predictors and the outcomes vary across team-level and facility- level units | Model 3 results + results that reveal the relationships between team-level predictors and the outcomes | Model 4 results + findings that shows if the associations between team-level predictors and the outcomes vary across facility-level units | Model 5 results + results that indicate the relationships between team-level predictors and the outcomes. |

Table 3: Estimated random intercept (co)variances introduced by VA facilities

| Components | Posterior Mean | L-95% HPD | U-95% HPD |
|---|---|---|---|
| PCRVU, PCRVU | 0.218 | 0.215 | 0.220 |
| PCRVU, Non-PCRVU | − 0.006 | − 0.009 | − 0.002 |
| Non-PCRVU, Non-PCRVU | 0.157 | 0.153 | 0.162 |

Table 4: Estimated random intercept (co)variances introduced by PCMH teams

| Components | Posterior Mean | L-95% HPD | U-95% HPD |
|---|---|---|---|
| PCRVU, PCRVU | 0.168 | 0.158 | 0.176 |
| PCRVU, Non-PCRVU | 0.035 | − 0.014 | 0.059 |
| Non-PCRVU, Non-PCRVU | 0.053 | 0.049 | 0.057 |

Table 5: Estimated patient-level residual (co)variances

| Components | Posterior Mean | L-95% HPD | U-95% HPD |
|---|---|---|---|
| PCRVU, PCRVU | 0.609 | 0.604 | 0.614 |
| PCRVU, Non-PCRVU | 0.316 | 0.298 | 0.330 |
| Non-PCRVU, Non-PCRVU | 0.787 | 0781 | 0.792 |

Table 6: Coefficient estimates from 3-level hierarchical model for joint PC and Non-PC workloads

|  | Model 1 | Model 2 | Model 3 | Model 4 | Model 5 | Model 6 |
|---|---|---|---|---|---|---|
| **Deterministic Effect** | | | | | | |
| Gender, Male | | 0.41″, 0.02″ | 0.43″, 0.03″ | 0.42″, 0.01″ | 0.42″, 0.02″ | 0.43″, 0.02″ |
| Age | | 1.02″, 1.04″ | 1.03, 1.03 | 1.03, 1.03 | 1.01, 1.02 | 1.02, 1.04 |
| Age × Age | | 0.92″, 0.94″ | 0.9′, 0.91′ | 0.91′, 0.93′ | 0.92′, 0.94′ | 0.92′, 0.93′ |
| Insurance, Yes | | 0.95′, 0.92 | 0.94′, 0.9 | 0.95′, 0.91 | 0.93′, 0.9 | 0.93′, 0.92 |
| LOS, Zero | | 1.07, 0.74″ | 1.06, 0.71″ | 1.08, 0.73″ | 1.07, 0.72″ | 1.07, 0.72″ |
| CAN Score | | 1.12″, 1.07″ | 1.08, 1.02 | 1.09, 1.03 | 1.08, 1.03 | 1.1, 1.02 |
| SQRT (CAN Score) | | 1.15′, 1.19″ | 1.1′, 1.12′ | 1.12′, 1.13′ | 1.11′, 1.12′ | 1.12′, 1.13′ |
| Priority (ref = 8) | | | | | | |
| 1 (disability > 50%) | | 0.96″, 1.25″ | 0.97″, 1.22″ | 0.96″, 1.23″ | 0.95″, 1.24″ | 0.95″, 1.23″ |
| 2 (disability 30%–40%) | | 1.02′, 1.32″ | 1.02′, 1.28′ | 1.03′, 1.29′ | 1.01′, 1.3′ | 1.01′, 1.28′ |
| 3 (disability 20–30%) | | 0.94′, 1.01″ | 0.92′, 1.04″ | 0.92′, 1.03″ | 0.93′, 1.04″ | 0.94′, 1.04″ |
| 4 (catastrophically dis.) | | 1.03″, 1.17″ | 1.04″, 1.14″ | 1.03′, 1.15″ | 1.05′, 1.15″ | 1.04′, 1.15″ |
| 5 (Medicaid) | | 1.05″, 1.03″ | 1.04″, 1.05″ | 1.05″, 1.04″ | 1.05″, 1.03″ | 1.05″, 1.04″ |
| 6 (Agent Orange, …) | | 1.06″, 1.34′ | 1.03″, 1.32″ | 1.03″, 1.33″ | 1.04″, 1.35″ | 1.05″, 1.35″ |
| 7 (below HUD) | | 1.09″, 1.1″ | 1.08″, 1.07″ | 1.09″, 1.07″ | 1.08″, 1.09″ | 1.08″, 1.1″ |
| ACC001–Infectious and Parasitic | | 1.07″, 1.22″ | 1.05″, 1.23″ | 1.04″, 1.24″ | 1.04″, 1.22″ | 1.05″, 1.22″ |
| ACC002–Malignant Neoplasm | | 1.04″, 1.33″ | 1.04″, 1.3″ | 1.03″, 1.31″ | 1.04″, 1.32″ | 1.03″, 1.31″ |
| ACC003–Benign/In Situ/Uncertain Neoplasm | | 1.07″, 1.65″ | 1.06″, 1.65″ | 1.06″, 1.64″ | 1.07″, 1.64″ | 1.06″, 1.64″ |
| ACC004–Diabetes | | 1.53″, 0.98′ | 1.52″, 0.97′ | 1.53″, 0.96′ | 1.53″, 0.97′ | 1.52″, 0.98′ |
| ACC005–Nutritional and Metabolic | | 1.18″, 1.02 | 1.19″, 1.03 | 1.2″, 1.02 | 1.2″, 1.04 | 1.19″, 1.03 |
| ACC006–Liver | | 1.13″, 1.04′ | 1.11″, 1.05′ | 1.12″, 1.05′ | 1.11″, 1.05′ | 1.11″, 1.04′ |
| ACC007–Gastrointestinal | | 1.09″, 1.13″ | 1.07″, 1.14″ | 1.07″, 1.14″ | 1.08″, 1.15″ | 1.08″, 1.14″ |
| ACC008–Musculoskeletal and Connective Tissue | | 1.18″, 1.27″ | 1.17″, 1.27″ | 1.16″, 1.28″ | 1.16″, 1.27″ | 1.17″, 1.26″ |
| ACC009–Hematological | | 1.09″, 1.05″ | 1.08″, 1.06″ | 1.07″, 1.06″ | 1.08″, 1.04″ | 1.08″, 1.05″ |
| ACC010–Cognitive Disorders | | 1, 1.12″ | 0.98, 1.1″ | 1, 1.11″ | 0.99, 1.12″ | 1.1, 1.12″ |
| ACC011–Substance Abuse | | 1.06″, 0.88″ | 1.06″, 0.9″ | 1.05″, 0.9″ | 1.05″, 0.89″ | 1.06″, 0.89″ |

Table 6—continued: Coefficient estimates from 3-level hierarchical model for joint PC and Non-PC workloads

|  | Model 1 | Model 2 | Model 3 | Model 4 | Model 5 | Model 6 |
|---|---|---|---|---|---|---|
| ACC012–Mental |  | 1.03′, 1.73″ | 1.04′, 1.7″ | 1.03′, 1.71″ | 1.04′, 1.71″ | 1.04′, 1.73″ |
| ACC013–Developmental Disability |  | 0.99, 1.24″ | 1.01, 1.23″ | 1.01, 1.22″ | 1, 1.22″ | 1.01, 1.24″ |
| ACC014–Neurological |  | 1.07″, 1.15″ | 1.06″, 1.14″ | 1.07″, 1.16″ | 1.07″, 1.15″ | 1.07″, 1.14″ |
| ACC015−Cardio-Respiratory Arrest |  | 1.07′, 1.02 | 1.03′, 1.04 | 1.05′, 1.03 | 1.05′, 1.03 | 1.06′, 1.03 |
| ACC016−Heart |  | 1.15″, 1.05′ | 1.14″, 1.06′ | 1.16″, 1.04′ | 1.15″, 1.05′ | 1.15″, 1.05′ |
| ACC017−Cerebrovascular |  | 1.05, 1.02 | 1.05, 1.03 | 1.04, 1.01 | 1.03, 0.99 | 1.04, 1.01 |
| ACC018−Vascular |  | 1.08″, 1.26″ | 1.1″, 1.26″ | 1.09″, 1.27″ | 1.11″, 1.27″ | 1.09″, 1.26″ |
| ACC019−Lung |  | 1.09″, 1.11″ | 1.07″, 1.12″ | 1.08″, 1.12″ | 1.08″, 1.11″ | 1.08″, 1.1″ |
| ACC020−Eyes |  | 1.08″, 1.12″ | 1.09″, 1.13″ | 1.09″, 1.14″ | 1.07″, 1.13″ | 1.07″, 1.11″ |
| ACC021−Ears, Nose, and Throat |  | 1.11″, 1.40″ | 1.12″, 1.38″ | 1.1″, 1.39″ | 1.12″, 1.39″ | 1.11″, 1.39″ |
| ACC022−Urinary System |  | 1.06″, 1.01 | 1.07″, 1.02 | 1.08″, 1.02 | 1.08″, 1.01 | 1.07″, 1.01 |
| ACC023−Genital System |  | 1.09″, 1.07″ | 1.09″, 1.04″ | 1.1″, 1.06″ | 1.11″, 1.05″ | 1.11″, 1.06″ |
| ACC025−Skin and Subcutaneous |  | 1.11″, 1.42″ | 1.13″, 1.43″ | 1.12″, 1.43″ | 1.13″, 1.41″ | 1.12″, 1.41″ |
| ACC026−Injury, Poisoning, Complications |  | 1.1″, 1.28″ | 1.11″, 1.29″ | 1.12″, 1.3″ | 1.12″, 1.29″ | 1.12″, 1.29″ |
| ACC027−Symptoms, Signs, and Ill-Defined Conditions |  | 1.17″, 1.45″ | 1.15″, 1.41″ | 1.16″, 1.42″ | 1.17″, 1.44″ | 1.16″, 1.43″ |
| ACC029−Transplants, Openings, Amputations |  | 0.9″, 1.01 | 0.94″, 0.98 | 0.92″, 0.99 | 0.93″, 1 | 0.93″, 1.01 |
| ACC030−Screening/History |  | 1.22″, 2.01″ | 1.23″, 1.98″ | 1.2″, 1.98″ | 1.2″, 2″ | 1.22″, 1.99″ |
| *Changed provider count* |  |  |  | 1.11″, 1.09″ | 1.04′, 1.03′ | 1.03′, 1.02′ |
| *Distance* (ref = Far) |  |  |  |  |  |  |
|   Middle |  |  |  |  |  | 0.89″, 0.87′ |
|   Near |  |  |  |  |  | 0.85″, 0.93′ |
| Diabetes × Liver |  | 1.02, 1.13′ | 1.03, 1.15′ | 1.03, 1.16′ | 1.01, 1.16′ | 1.01, 1.14′ |
| Diabetes × Cardio-Respiratory Arrest |  | 1.12′, 1.11″ | 1.1′, 1.13″ | 1.13′, 1.12″ | 1.15′, 1.14″ | 1.13′, 1.14″ |
| Diabetes × Heart |  | 1.03, 1.1″ | 1.04, 1.12″ | 1.03, 1.11″ | 1.01, 1.11″ | 1.01, 1.11″ |
| Diabetes × Cerebrovascular |  | 1.07′, 1.17′ | 1.06′, 1.14′ | 1.06′, 1.15′ | 1.06′, 1.16′ | 1.06′, 1.15′ |
| Diabetes × Urinary System |  | 1.04, 1.12′ | 1.06, 1.1′ | 1.05, 1.1′ | 1.07, 1.13′ | 1.06, 1.13′ |

Table 6—continued: Coefficient estimates from 3-level hierarchical model for joint PC and Non-PC workloads

|  | Model 1 | Model 2 | Model 3 | Model 4 | Model 5 | Model 6 |
|---|---|---|---|---|---|---|
| Diabetes × Transplants, Openings, Amputations |  | 1.08′, 1.09 | 1.09′, 1.07 | 1.09′, 1.08 | 1.11′, 1.09 | 1.1′, 1.09 |
| Substance Abuse × Mental |  | 1.04″, 1.20″ | 1.03″, 1.21″ | 1.04″, 1.21″ | 1.05″, 1.22″ | 1.04″, 1.21″ |
| Heart × Cerebrovascular |  | 1.12′, 1.14″ | 1.09′, 1.13″ | 1.1′, 1.15″ | 1.11′, 1.16″ | 1.11′, 1.16″ |
| Heart × Vascular |  | 1.06, 1.04′ | 1.07, 1.05′ | 1.05, 1.05′ | 1.05, 1.06′ | 1.04, 1.06′ |
| Cerebrovascular × Vascular |  | 1.01, 1.12″ | 1.03, 1.13″ | 1.02, 1.14″ | 1.03, 1.14″ | 1.03, 1.12″ |
| Male × Diabetes |  | 1.06′, 1.12′ | 1.05′, 1.14′ | 1.04′, 1.14′ | 1.07′, 1.11′ | 1.05′, 1.13′ |
| Male × Neurological |  | 1.08″, 1.11′ | 1.09″, 1.13′ | 1.1″, 1.12′ | 1.13″, 1.12′ | 1.1″, 1.12′ |
| Age × Heart |  | 1.11″, 1.21′ | 1.09″, 1.19′ | 1.09″, 1.2′ | 1.08″, 1.22′ | 1.08″, 1.2′ |
| Age × Nutritional and Metabolic |  | 1.14′, 1.07″ | 1.15′, 1.09″ | 1.14′, 1.08″ | 1.14′, 1.1″ | 1.16′, 1.09″ |
| Age × Gastrointestinal |  | 1.05′, 1.1′ | 1.07′, 1.12′ | 1.06′, 1.12′ | 1.08′, 1.09′ | 1.08′, 1.1′ |
| Priority 4 × Neurological |  | 1.13′, 1.17″ | 1.14′, 1.17″ | 1.11′, 1.16″ | 1.15′, 1.16″ | 1.14′, 1.16″ |
| Priority 6 × Cardio-Respiratory Arrest |  | 1.14″, 1.06′ | 1.14″, 1.07′ | 1.13″, 1.07′ | 1.11″, 1.09′ | 1.11″, 1.07′ |
| **Variance Component** |  |  |  |  |  |  |
| Residual | 0.609′, 0.79′ | 0.446′, 0.55′ | 0.357′, 0.46′ | 0.352′, 0.44′ | 0.259′, 0.43′ | 0.255′, 0.43′ |
| Intercept (team) | 0.168′, 0.05′ | 0.093′, 0.04′ | 0.076′, 0.04′ | 0.064′, 0.04′ | 0.059′, 0.03′ | 0.054′, 0.03′ |
| Intercept (facility) | 0.218′, 0.16′ | 0.125′, 0.1′ | 0.106′, 0.08′ | 0.091′, 0.08′ | 0.085′, 0.07′ | 0.083′, 0.05′ |
| Slope (age: team) |  |  | 0.088′, 0.09′ | 0.081′, 0.09′ | 0.075′, 0.1′ | 0.073′, 0.08′ |
| Slope (age^2: team) |  |  | 0.042′, 0.06 | 0.047, 0.07′ | 0.053, 0.06′ | 0.056′, 0.06′ |
| Slope (CAN Score: team) |  |  | 0.078′, 0.09′ | 0.072′, 0.1′ | 0.069′, 0.09′ | 0.066′, 0.09′ |
| Slope (CAN Score^(0.5): team) |  |  | 0.037, 0.05′ | 0.042′, 0.04′ | 0.038′, 0.04 | 0.041, 0.05′ |
| Slope (insurance: facility) |  |  | 0.051′, 0.06′ | 0.047′, 0.07′ | 0.049′, 0.06′ | 0.045′, 0.07′ |
| Slope (changed provider count: facility) |  |  |  |  | 0.053′, 0.06′ | 0.046′, 0.05′ |
| **Model Fit** |  |  |  |  |  |  |
| DIC | 461019.6 | 227245.2 | 225469.7 | 225411.4 | 225351.3 | 225337.8 |

Table 7: Goodness-of-fit values for the two scenarios

| Facility | Team | Patient | DIC |
|---|---|---|---|
| 2 | 2 | 2 | 225337.8 – 227448.1 |
| 2 | 2 | 1 | 225491.7 – 225494.1 |
| 2 | 1 | 2 | 225401.1 – 225396.9 |
| 2 | 1 | 1 | 225582.5 – 225580.3 |
| 1 | 2 | 2 | 225378.5 – 225375.7 |
| 1 | 2 | 1 | 225444.9 – 225441.2 |
| 1 | 1 | 2 | 225457.8 – 225460.5 |
| 1 | 1 | 1 | 225550.7 – 225554.0 |

Table 8: Summary statistics for the range of joint correct intervals

|  | Multivariate | | Univariate | |
|---|---|---|---|---|
|  | PC | Non-PC | PC | Non-PC |
| Mean | 0.431 | 1.023 | 0.514 | 1.083 |
| Median | 0.381 | 0.977 | 0.439 | 1.058 |